\documentclass[11pt,letterpaper]{article}
\usepackage{systeme}
\usepackage[utf8]{inputenc}
\usepackage{multirow}
\usepackage{booktabs}
\usepackage[english]{babel}
\usepackage{amsmath}
\usepackage{amsfonts}
\usepackage{amssymb}
\usepackage{lmodern}
\usepackage{graphicx}
\usepackage[small,bf]{caption}
\usepackage[margin=2.5cm]{geometry}
\usepackage{framed}
\usepackage{color}
\usepackage{wrapfig}\definecolor{shadecolor}{RGB}{220,220,220}
\usepackage{float}
\usepackage{array}
\usepackage{caption}
\usepackage{microtype}
\usepackage{titlesec}
\usepackage{enumitem}
\usepackage{tcolorbox}
\titleformat{\section}
  {\large\bfseries}
  {\thesection}{1em}{}
\titleformat{\subsection}
  {\normalsize\bfseries}
  {\thesubsection}{1em}{}
\titlespacing*{\section}{0pt}{16pt}{6pt}
\titlespacing*{\subsection}{0pt}{12pt}{4pt}

\title{\textbf{Gravity, Entropy, and the Second Law\\
 A Pedagogical Perspective}}

\begin{document}
\author{Jorge Pinochet$^{1}$, Giorgio Sonnino$^{2}$\\ \\
\small{$^{1}$\textit{Department of Physics, Centro de Desarrollo de Investigación CEDI-UMCE,}}\\
  \small{\textit{Universidad Metropolitana de Ciencias de la Educación,}}\\
 \small{\textit{Av. José Pedro Alessandri 774, Ñuñoa, Santiago, Chile.}}\\
\small{$^{2}$\textit{Université Libre de Bruxelles (U.L.B.) \&}}\\
\small\textit{International SOLVAY Institutes for Physics and Chemistry}\\
\small{\textit{Campus de la Plaine C.P. 224, Bvd du Triomphe, 1050 Brussels, Belgium.}}\\
\small{e-mail: jorge.pinochet@umce.cl\ ; e-mail: giorgio.sonnino@ulb.be}}

\date{}
\maketitle

\begin{center}\rule{0.9\textwidth}{0.1mm} \end{center}
\begin{abstract}
\noindent An interesting question for discussion in university physics courses is whether gravity can conflict with the second law of thermodynamics. In this work, we address this question by comparing the behavior of an ideal gas in two conceptually different regimes: one in which gravitational effects can be neglected and another in which self-gravity becomes important. In the first case, free expansion provides the familiar example in which an increase in the volume accessible to the gas leads to an increase in entropy. Self-gravitating systems behave differently: gravitational contraction can reduce the volume occupied by matter while releasing energy and increasing the entropy of the surroundings. This illustrates why the usual intuition that increasing entropy necessarily corresponds to increasing spatial dispersion cannot be applied directly to self-gravitating systems. Using elementary thermodynamic arguments and a minimal amount of mathematics, we discuss this apparent tension and show how entropy must be considered for the complete system, including emitted radiation. The discussion is finally extended to black holes and Hawking radiation, where the appropriate framework is the generalized second law of thermodynamics. The analysis is intended to provide a pedagogical bridge between elementary thermodynamics and some of the distinctive thermodynamic properties of gravitating systems.


\begin{center}\rule{0.9\textwidth}{0.1mm} \end{center}
\end{abstract}

\section{Introduction}\label{I}

Let us imagine a small hermetically sealed container filled with an ideal gas. The container is divided into two regions separated by a barrier. When the barrier is removed, the gas spontaneously expands into the initially empty region. After reaching equilibrium, it occupies the entire available volume and has a larger thermodynamic entropy, as required by the second law of thermodynamics \cite{fermi}.\\

Now imagine that we gradually increase both the size of the container and the mass of the gas it contains. At some point, the effects of self-gravity can no longer be neglected. The behavior of the gas then changes dramatically: instead of expanding, the matter tends to contract and form increasingly concentrated structures. This behavior not only appears to conflict with the second law of thermodynamics, but also with the familiar behavior of an ideal gas in the absence of significant gravitational effects, which tends to expand and distribute itself throughout the available volume. This naturally leads to a fundamental question: Can gravity violate the second law of thermodynamics? Posing this question in a university course on thermodynamics or statistical physics offers an opportunity to stimulate debate and reflect on the physical meaning of entropy.\\

The relationship between gravity and entropy is rarely discussed in university textbooks on general physics and thermodynamics \cite{pinochet1}. In this work, we address this question by comparing two conceptually different situations: an ideal gas whose self-gravity can be neglected and a gas in which gravitational effects become important. We begin with free expansion and then analyze gravitational contraction, showing how the apparent decrease in entropy associated with the concentration of matter is resolved by considering the complete entropy budget. Finally, we extend the discussion to the extreme case of black holes and Bekenstein-Hawking entropy.

\section{''Turning Off'' Gravity: Entropy in the Free Expansion}\label{GFE}

Let us begin by analyzing the free expansion of an ideal gas under the assumption that gravity has been ``turned off,'' which is effectively what all general physics textbooks do, even if they do not state it explicitly. This approximation is valid provided that the mass of the gas is not large enough for the gravitational attraction between particles to become significant. Since this topic is treated in great detail in most thermodynamics textbooks, we will focus only on the central aspects of the process.\\

\begin{figure}[H]
\centering
\includegraphics[width=0.5\textwidth]{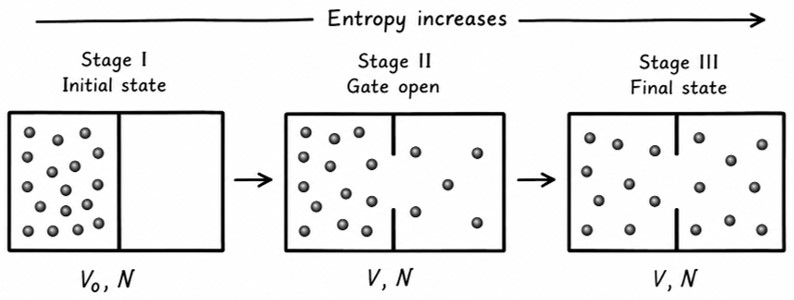}
\caption{Free expansion of an ideal gas. The entropy increases as the gas expands.}\label{Fig1}
\end{figure}

Figure~\ref{Fig1} illustrates the free expansion, which we have divided into three stages. Recall that this is an irreversible process in which a gas composed of a large number of noninteracting point particles spontaneously expands into an evacuated and isolated container \cite{reif}. 

As the gas occupies an increasingly larger volume, its distribution becomes more uniform and its entropy increases. This greater uniformity is a useful way to visualize the increase in entropy, although it does not constitute its definition. In stage I, the ideal gas is confined to the left half of a closed and isolated container divided in two by a barrier. When the barrier is opened in stage II, the gas gradually expands into the right half, which is initially empty. Finally, in stage III, the free expansion causes the gas to occupy the entire available volume.\\ 

Since there is no exchange of energy with the external environment, this process is adiabatic. More precisely, the container is isolated: no heat or work crosses its boundary. For an ideal gas undergoing free expansion, the internal energy therefore remains constant, and because the internal energy of an ideal gas depends only on temperature, the initial and final equilibrium temperatures are equal. Let us now examine how the entropy changes between stages I and III. To do so, consider the well-known relation \cite{fermi}
\begin{equation}\label{GFE1}
\Delta S_{\mathrm{FE}} = N k_B \ln \left(\frac{V}{V_0}\right),
\end{equation}
where $\Delta S_{\mathrm{FE}}$ is the entropy variation during the free expansion (FE). In addition, $k_B = 1.38 \times 10^{-23}~\mathrm{kg \cdot m^2 \cdot K^{-1}}$ is Boltzmann's constant, $V$ is the final volume of the gas, $V_0$ is its initial volume, and $N$ is the number of particles (atoms), which is typically of the order of Avogadro's number, $N \sim 10^{23}$. Equation~(1), found in any general physics textbook, gives the entropy difference between the initial and final equilibrium states. Although the actual free expansion is irreversible, entropy is a state function, so the same difference may be calculated by imagining any convenient reversible path connecting those two states. The equation shows that the final entropy is larger because a greater volume is accessible to the gas particles. Indeed, since the gas volume increases, we have $V > V_0$, which implies that
\begin{equation}\label{GFE2}
\Delta S_{\mathrm{FE}} > 0
\end{equation}
which is consistent with the second law of thermodynamics. From a microscopic point of view, this result is explained as an evolution toward the most probable state. It is far more probable for the particles to occupy the entire container homogeneously than to remain confined within a reduced volume \cite{fermi}, \cite{rh}. Once the gas has relaxed to equilibrium throughout the available volume, its macroscopic entropy no longer changes systematically. In statistical mechanics, this equilibrium macrostate corresponds overwhelmingly to the largest region of accessible phase space; microscopic fluctuations nevertheless continue. The fundamental characteristic of this state is that, aside from small statistical fluctuations, the gas no longer undergoes macroscopically observable changes. Under these conditions, we say that the gas is in a state of thermal equilibrium \cite{reif}. In Section~\ref{EGC} we return to Eq.~(\ref{GFE1}) and apply it to an ideal gas subjected to its own self-gravity, where we examine what happens to the entropy variation.

\section{"Turning On" Gravity: Entropy in the Gravitational Contraction}\label{EGC}

Let us now imagine gradually increasing the size of the container illustrated in Figure~\ref{Fig1}, while simultaneously increasing the mass of the ideal gas it contains. At some point, when the container reaches astronomical dimensions and $N$ greatly exceeds Avogadro's number, the effects of self-gravity will begin to become important. Consequently, we will no longer be able to ignore them, as we did in the previous section and as is commonly done in physics and thermodynamics textbooks.

\begin{figure}[H]
\centering
\includegraphics[width=0.5\textwidth]{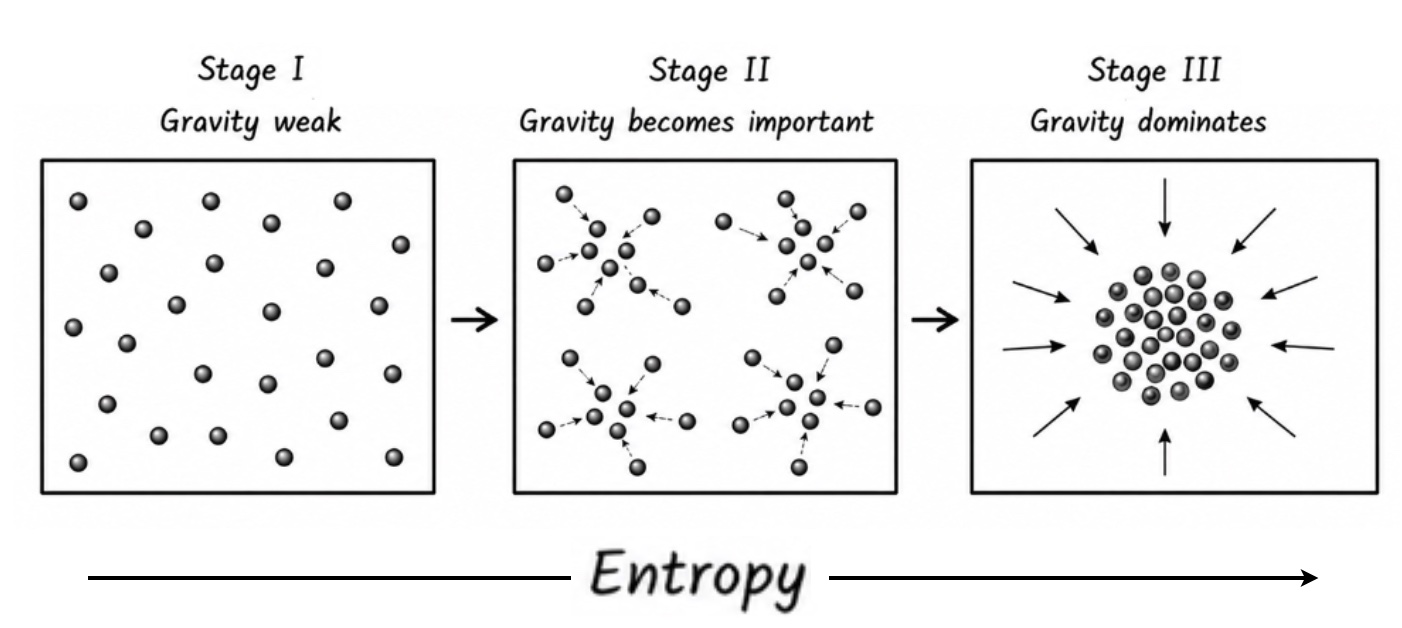}
\caption{Gravitational contraction evolves in the opposite direction to the free expansion.}\label{Fig2}
\end{figure}

The situation is illustrated in Figure~\ref{Fig2}, which presents an evolution opposite to that described in the previous section: whereas free expansion increases both the volume of the gas and the uniformity of the particle distribution, we now have a process of gravitational contraction (GC), which reduces both the volume and the uniformity of the gas. To analyze the process, let us return to Eq.~(\ref{GFE1}). If we naively apply only the volume term in Eq.~(\ref{GFE1}), keeping the other thermodynamic variables implicit and inserting $V < V_0$, we obtain a result that at first sight appears puzzling. This step is intentionally heuristic: Eq.~(\ref{GFE1}) was derived for an ideal gas between equilibrium states at the same temperature and without long-range gravitational interactions, so it cannot by itself describe a self-gravitating contraction.
\begin{equation}\label{EGC1}
\Delta S_{\mathrm{GC}} < 0
\end{equation}
This result is opposite to that obtained for free expansion and seems to contradict the second law of thermodynamics, since it suggests a decrease in entropy during the process of gravitational contraction.\\

The apparent contradiction is a warning that Eq.~(\ref{EGC1}) does not describe the entire process. In particular, self-gravity changes the energy balance and the temperature, and a contracting gravitating system generally exchanges energy internally between gravitational and kinetic degrees of freedom and may also produce radiation. Thus the entropy bookkeeping must include all relevant components and not only a ''volume entropy'' of the gas. In other words, the reduction in the entropy of the gas described by Eq.~(\ref{EGC1}) must be accompanied by a compensating increase in entropy in some component of the system unrelated to the decrease in volume. In the next section we will see where this missing entropy is hidden and how the second law is preserved during the process of gravitational contraction.

\section{Recovering the missing entropy}\label{RmE}

To identify the entropy apparently lost during gravitational contraction, we must consider an important result from the physics of self-gravitating systems known as the \textit{virial theorem}~\cite{reif}. According to the theorem, in an approximately stable self-gravitating system, the average kinetic energy $\langle K \rangle$ and the average gravitational potential energy $\langle U \rangle$ are related through $2\langle K \rangle + \langle U \rangle = 0$, or equivalently, $\langle K \rangle = -\langle U \rangle/2$, which implies that the average total energy of the system is $\langle E \rangle = \langle K \rangle + \langle U \rangle = \langle U \rangle/2 = -\langle K \rangle$. We see that the total energy is negative, which is a distinctive characteristic of gravitationally bound systems \cite{reif}. The theorem should not be interpreted as a direct conversion between potential and kinetic energy, but rather as a relation satisfied by a bound system after suitable time averaging (or by a sequence of approximately virialized states). This qualification is pedagogically important: during a rapid collapse the system need not satisfy the virial relation instantaneously.\\ 

Let us now analyze Eq.~(\ref{EGC1}) in light of the virial theorem. If the system undergoes a slow contraction, we may compare two successive states and write approximately \cite{pinochet2}
\begin{equation}\label{EGC2}
\Delta K \simeq -\frac{\Delta U}{2}
\end{equation}
According to this relation, when the gas contracts, the gravitational potential energy decreases, that is, it becomes more negative ($\Delta U < 0$), producing a compensating increase in the kinetic energy of the particles ($\Delta K > 0$). Since, in an ideal gas, the average kinetic energy is proportional to the absolute temperature, gravitational contraction naturally tends to heat the gas. If the gas can radiate, this increase in temperature promotes the emission of electromagnetic radiation. In the sealed-container thought experiment, the radiation does not leave the total isolated system: it propagates through, and may eventually be absorbed within, the container. In an astrophysical setting, the analogous radiation can escape to the surrounding space. This idea is fundamental for understanding why gravity does not violate the second law of thermodynamics. The radiation emitted by the system transports not only energy, but also entropy. For a schematic entropy budget, let us denote by $\Delta S_{\mathrm{Vol}}$ the negative contribution suggested by the reduction in the characteristic gas volume and by $\Delta S_{\mathrm{Rad}}$ the entropy carried by the radiation field (see Figure~\ref{Fig2}). These symbols are pedagogical bookkeeping devices rather than a complete thermodynamic decomposition, because in a self-gravitating system, volume, temperature, density profile, and gravitational energy are coupled. Under these conditions, the total entropy variation during the contraction process can be schematically written as
\begin{equation}\label{EGC3}
\Delta S_{\mathrm{GC}} =
\Delta S_{\mathrm{Vol}} +
\Delta S_{\mathrm{Rad}}.
\end{equation}
For the total isolated system, the second law requires the total entropy not to decrease. Within the simplified two-term bookkeeping adopted here, this means that the positive entropy contribution associated with radiation must compensate for the negative volume contribution. That is, the following condition must hold:
\begin{equation}\label{EGC4}
\lvert \Delta S_{\mathrm{Vol}} \rvert < \Delta S_{\mathrm{Rad}},
\end{equation}
so that, ultimately,
\begin{equation}\label{EGC5}
\Delta S_{\mathrm{GC}} > 0.
\end{equation}
This inequality tells us that gravity can produce local order while the global entropy continues to increase. A rigorous demonstration for a specified model requires a more detailed treatment of the gas, radiation field, boundary conditions, and gravitational interaction, and therefore lies beyond the scope of this work. Accordingly, Eq.~(\ref{EGC4}) should be read as the condition imposed by the second law within our simplified entropy budget, not as a general theorem derived solely from the virial relation.\\

\begin{figure}[H]
\centering
\includegraphics[width=0.7\textwidth]{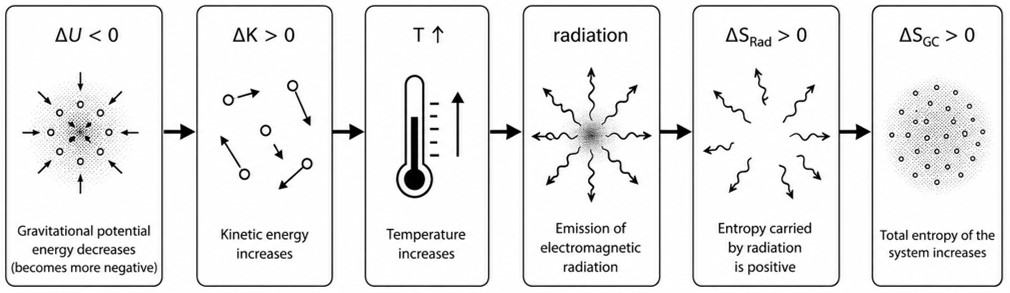}
\caption{Schematic summary of the complete physical process occurring during gravitational contraction.}\label{Fig3}
\end{figure}

The physical discussion nevertheless explains where the entropy omitted by the naive volume-only argument can reside. Although gravitational contraction reduces the volume and favors the formation of increasingly concentrated structures, the highly entropic radiation emitted during the process disperses throughout the surrounding space. Consequently, the entropy increase associated with the radiation more than compensates for the entropy decrease associated with the contraction of the gas. The final result is a net increase in the total entropy. The central physical argument discussed in this section may be summarized schematically as
\begin{equation}\label{scheme}
\Delta U < 0
\rightarrow
\Delta K > 0
\rightarrow
T \uparrow
\rightarrow
\text{radiation}
\rightarrow
\Delta S_{\mathrm{Rad}} > 0
\rightarrow
\Delta S_{\mathrm{GC}} > 0.
\end{equation}
That is, the contraction reduces the gravitational potential energy, which is compensated by an increase in the kinetic energy and temperature of the gas. When radiative processes are available, the resulting heating produces radiation, which provides an important positive contribution to the entropy budget. The complete second-law statement always concerns the entropy of the whole isolated system (or, for a non-isolated subsystem, the entropy production together with entropy transferred across its boundary). Figure~\ref{Fig3} graphically illustrates the complete process, which preserves the second law of thermodynamics.

From a microscopic point of view, the comparison with free expansion must therefore be made with care. Long-range attractive gravity makes a nearly homogeneous distribution susceptible, under appropriate conditions, to gravitational instability: a small overdensity attracts additional matter and can grow rather than being erased. This is the qualitative idea behind gravitational clumping (and, in more advanced treatments, the Jeans instability). Thus "uniformity" is not a universal proxy for high entropy once gravity is dynamically important. Self-gravitating systems also possess unusual thermodynamic properties; for example, bound configurations can exhibit negative heat capacity, so the intuition developed for short-range laboratory gases cannot be transferred without qualification.

\section{Gravity Taken to the Extreme: The Bekenstein-Hawking Entropy}\label{BH}

The ideas developed in the previous sections show that gravity does not violate the second law of thermodynamics, even when it produces increasingly concentrated structures. However, the relationship between gravity and entropy reaches its most extreme expression in the physics of black holes, making it important to analyze this scenario to construct as complete a picture as possible of the problem under consideration. Black holes are the most extreme prediction of general relativity, Einstein's theory of gravity. The gravitational field generated by a black hole is so intense that nothing, not even light, can escape from its interior. For a long time, it was believed that, due to this apparent inability to emit radiation, black holes could not possess either temperature or entropy. However, in 1973, the Israeli physicist Jacob Bekenstein proposed that black holes do possess entropy. Shortly afterward, the British theoretical physicist Stephen Hawking obtained the exact mathematical expression, now known as the \textit{Bekenstein-Hawking entropy}~\cite{B,H1,H2,pinochet3}:
\begin{equation}\label{BH1}
S_{\mathrm{BH}} =
\frac{4 \pi k_B G M_{\mathrm{BH}}^2}{\hbar c},
\end{equation}
where $G = 6.67 \times 10^{-11}~\mathrm{m^3\,kg^{-1}\,s^{-2}}$ is Newton's gravitational constant, $c = 3 \times 10^8~\mathrm{m \cdot s^{-1}}$ is the speed of light in vacuum, $\hbar = h/2\pi = 1.05 \times 10^{-34}~\mathrm{J \cdot s}$ is the reduced Planck constant, and $M_{\mathrm{BH}}$ is the mass of the black hole.\\

To gain an idea of the magnitudes involved, let us evaluate $S_{\mathrm{BH}}$ for a black hole with a mass of the order of one solar mass. This is an illustrative reference scale rather than a strict observational lower bound; known astrophysical black holes span a range of masses, while hypothetical primordial black holes could in principle be much lighter. We use $M_{\odot} = 1.99 \times 10^{30}~\mathrm{kg}$. Taking $M_{\mathrm{BH}} = M_{\odot}$ in Eq.~(9), and substituting the numerical values of the physical constants, we obtain $S_{\mathrm{BH}} \sim 10^{54}~\mathrm{J \cdot K^{-1}}$. This result is remarkable because it reveals that black holes possess colossal amounts of entropy. In fact, it can be shown that the maximum entropy that can exist within a given region is precisely the entropy that would be generated by a black hole occupying that region \cite{pinochet2}. Shortly afterward, Hawking also demonstrated that black holes possess an associated absolute temperature, inversely proportional to their mass, given by the equation \cite{H1}
\begin{equation}\label{BH2}
T_H =\frac{\hbar c^3}{8 \pi k_B G M_{\mathrm{BH}}},
\end{equation}
In semiclassical gravity, this temperature is associated with an approximately thermal flux known as \textit{Hawking radiation}. It is important not to picture the black hole as an ordinary hot material surface: Hawking radiation is not emitted by a hot surface, but is instead a quantum effect associated with the event horizon. However, due to Einstein's mass-energy equivalence, $E = mc^2$, the emitted energy gradually reduces the mass of the black hole through a process known as evaporation~\cite{H1}.

\begin{figure}[H]
\centering
\includegraphics[width=0.5\textwidth]{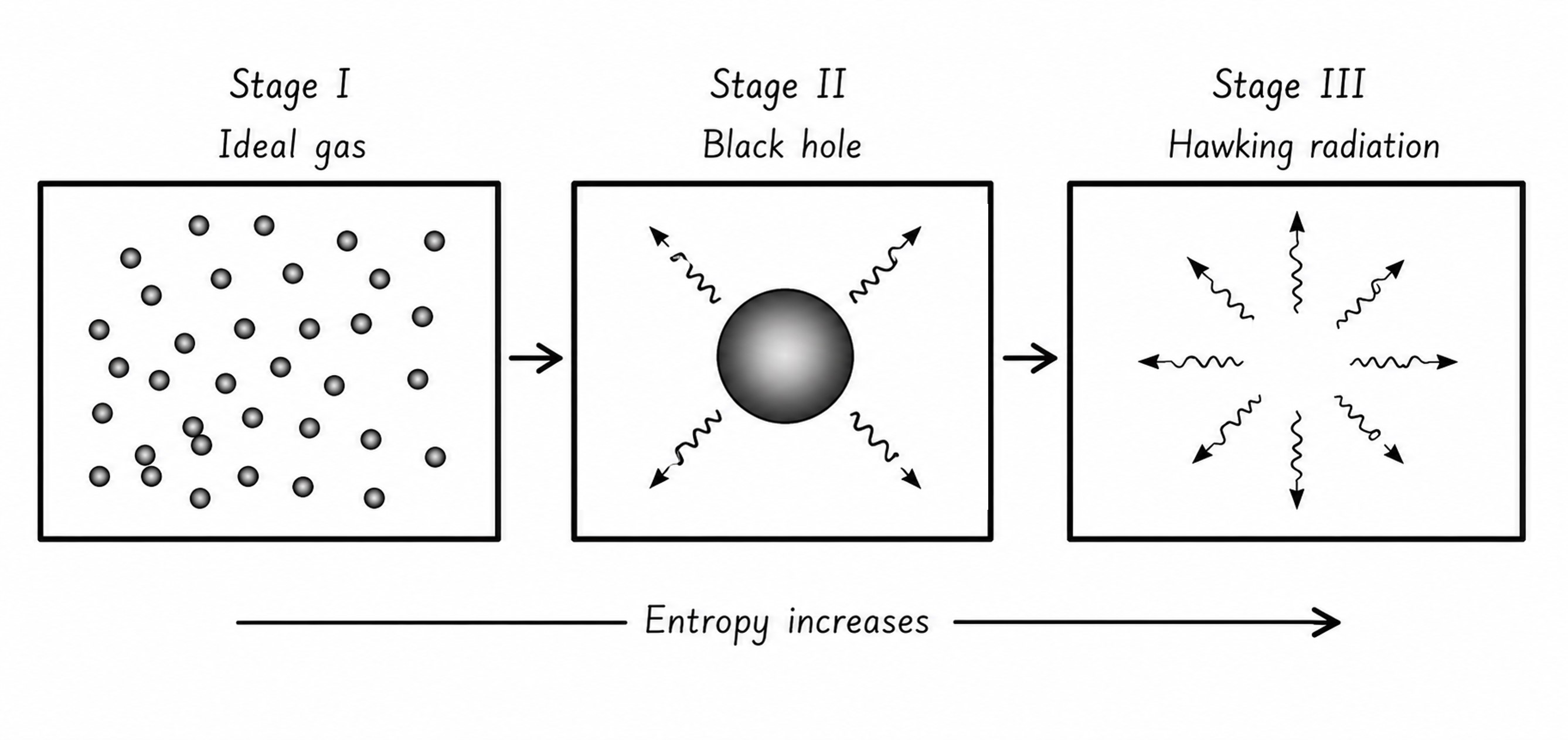}
\caption{Gravitational contraction of a very large mass of ideal gas.}\label{Fig4}
\end{figure}

These ideas naturally lead to the following question: how does the gas evolve during gravitational contraction when the mass is extremely large? In general terms, and without entering into the complex astrophysical processes governing the evolution of large gas masses under their own gravity, the process illustrated in Figure~\ref{Fig4} may be summarized as follows:

\begin{enumerate}[label=\Roman*.]
\item The gas mass contracts and heats up, emitting thermal radiation.
\item A black hole forms, which, for simplicity, we may assume absorbs the entire contents of the container, including the radiation emitted during stage I.
\item The black hole completely evaporates, leaving the container filled only with Hawking radiation.
\end{enumerate}

The appropriate second-law statement in the presence of a black hole is the \textit{generalized second law}: the sum of the entropy outside the black hole and the Bekenstein-Hawking entropy of the horizon does not decrease. Thus one must include the horizon entropy while the black hole exists, and the entropy of the emitted radiation during evaporation. Indeed, from the ideas developed in the previous sections, we know that the evolution from stage I to II (Figure~\ref{Fig4}) involves a reduction in the gas volume and an increase in the total entropy. We may summarize this conclusion as
\begin{equation}\label{BH3}
\Delta S_{(\mathrm{I} \rightarrow \mathrm{II})} > 0
\end{equation}
It can also be shown that the evolution from stage II to stage III leads to an increase in entropy. If $S_{\mathrm{HR}}$ denotes the entropy of the Hawking radiation after complete evaporation, a commonly used idealized, quasistatic blackbody estimate gives
\begin{equation}\label{BH4}
S_{\mathrm{HR}} \simeq \frac{4 S_{\mathrm{BH}}}{3}
\end{equation}
so that
\begin{equation}\label{BH5}
\Delta S_{(\mathrm{II} \rightarrow \mathrm{III})}=S_{\mathrm{HR}} - S_{\mathrm{BH}}\simeq\frac{4 S_{\mathrm{BH}}}{3} - S_{\mathrm{BH}}=\frac{S_{\mathrm{BH}}}{3}> 0
\end{equation}
Combining the two previous results, we finally obtain
\begin{equation}\label{BH6}
\Delta S_{(\mathrm{I} \rightarrow \mathrm{II})}+\Delta S_{(\mathrm{II} \rightarrow \mathrm{III})}> 0
\end{equation}
Within this idealized estimate, the generalized second law is once again satisfied. This result depends on the approximations used and should not be regarded as an exact result in all cases. The three stages leading to the complete evaporation of the black hole have never been observed because, among other reasons, the process would require a timescale far greater than the current age of the Universe. Therefore, the process illustrated in Figure~\ref{Fig4} remains theoretical. Hawking radiation is a robust prediction of semiclassical gravity, although it has not yet been directly detected from an astrophysical black hole \cite{matson, kfc}. The broader lesson for the present discussion is not that "uniformity'' must always prevail, but that gravitational clumping and increasing entropy are compatible once all relevant entropy contributions - matter, radiation, and, when present, black-hole horizon entropy - are included.

\section{Educational considerations}\label{EC}

The discussion presented in this work offers valuable learning opportunities, as it connects thermodynamics, statistical physics, gravitation, and astrophysics. Its main educational value does not lie in introducing a new formalism, but in using a familiar concept---entropy---to explore situations in which some of our usual physical intuitions become inadequate.\\

The contrast between free expansion and gravitational contraction provides a simple starting point for classroom discussion. In the first case, the increase in entropy is naturally associated with an increase in the volume accessible to the gas. In the second, gravity produces increasingly concentrated structures, which apparently suggests the opposite behavior. This contrast offers an opportunity to question the identification of entropy with disorder or spatial uniformity, and to emphasize that entropy is a thermodynamic quantity whose interpretation depends on the complete physical system being considered.\\

The discussion can also be used to help students distinguish between a local decrease in entropy and the change in entropy of the complete system. In a self-gravitating system, the contraction of matter can lead to a local decrease in entropy, but this does not represent the complete entropy budget: heating and radiation generate additional contributions that must be included to determine the total change in entropy. Extending the argument to black holes then provides a natural transition from classical thermodynamics to the generalized second law and Bekenstein-Hawking entropy.\\

These ideas can be explored through simple questions that stimulate discussion before introducing the mathematical description. For example:

\begin{itemize}
    \item Why does a gas normally expand while a sufficiently massive self-gravitating gas can contract?
    \item Does a more concentrated distribution necessarily have lower entropy?
    \item Where does the entropy go during gravitational contraction?
    \item What contribution must be included in the entropy budget when a black hole forms?
\end{itemize}

The equations presented in this work can then be used to support the conceptual arguments rather than becoming the main focus of the activity.

\section{Conclusions}\label{C}

We have seen that, using simple arguments, intuitive ideas, and only a minimal amount of mathematics, it is possible to understand why gravity does not violate the second law of thermodynamics, even when it produces increasingly concentrated structures. The key pedagogical point is that entropy should not be identified simply with spatial uniformity. For a self-gravitating system, the complete entropy budget must be considered, including the contributions from matter and radiation and, in the black-hole regime, the Bekenstein-Hawking entropy. Eq.~(\ref{BH6}) illustrates the consistency of the process with the second law of thermodynamics.\\

It is important to emphasize that the complete evaporation of a black hole has never been observed. As noted above, observing the full process would require a timescale vastly exceeding the present age of the Universe. Consequently, this part of our discussion remains theoretical, although it is supported by well-established theoretical arguments \cite{sonnino}.\\

We hope that this work will contribute to stimulating discussions on the relationship between gravity, entropy, and statistical physics in university courses on thermodynamics and modern physics.

\end{document}